\documentclass[a4paper,conference]{IEEEtran}
\usepackage[T1]{fontenc}
\usepackage[utf8]{inputenc}
\usepackage{floatrow} % Centers the content of figures.
\usepackage{subfig}
\usepackage{pgfplots}
\usepackage{tikz}
\usepackage{paralist}
\usepackage{amsmath}
\usepackage{amssymb}
\usepackage{algorithm}
\usepackage{algorithmic}

\usetikzlibrary{positioning}
\usetikzlibrary{topaths}

\tikzstyle{node} = [circle, draw, minimum size = 0.6 cm]

\newcommand{\SSC}[1]{\{#1\}}
\newcommand{\FSSC}{\Omega}
\newcommand{\elabel}[2]{(#1, #2)}
\newcommand{\nlabel}[3]{(#1, #2, #3)}
\newcommand{\nulledge}{e_\emptyset}

\title{Adapted and Constrained Dijkstra\\for Elastic Optical Networks}

\author{\IEEEauthorblockN{Ireneusz Szcześniak}
  \IEEEauthorblockA{
    AGH University of Science and Technology\\
    Department of Telecommunications\\
    al.~Mickiewicza 30\\
    30-059 Krakow\\
    Poland}
  \and
  \IEEEauthorblockN{Bożena Woźna-Szcześniak}
  \IEEEauthorblockA{
    Jan Długosz University in Częstochowa\\
    Faculty of Mathematics and Natural Sciences\\
    Institute of Mathematics and Computer Science\\
    al.~Armii Krajowej 13/15\\
    42-200 Częstochowa\\
    Poland}
}

\begin{document}

\maketitle

\begin{abstract}
  We present an optimal and efficient algorithm for finding a shortest
  path in an elastic optical network.  The algorithm is an adaptation
  of the Dijkstra shortest path algorithm, where we take into account
  the spectrum continuity and contiguity constraints, and a limit on
  the path length.  The adaptation redefines the node label in the
  Dijkstra algorithm, allows for revisiting nodes even at a higher cost
  for different slices, avoids loops, and prunes worse labels.  The
  algorithm is generic and agnostic of a specific spectrum allocation
  policy, as it finds the largest set of available slices from which
  slices can be allocated in any way.  We describe and motivate the
  algorithm design, and point to our freely-available implementation
  using the Boost Graph Library.  We carried out 8100 simulation runs
  for large, random and realistic networks, and found that the
  probability of establishing a connection using the proposed
  algorithm can be even twice as large as the probability of
  establishing a connection using the edge-disjoint shortest paths,
  and the Yen $K$ shortest paths.
\end{abstract}


%%%%%%%%%%%%%%%%%%%%%%%%%%%%%%%%%%%%%%%%%%%%%%%%%%%%%%%%%%%%%%%%%%%%%%%%%%%

\section{Introduction}
\label{introduction}

% On EONs briefly.

Elastic optical networks (EON), a.k.a., the flex-grid networks, are
considered the successor of wavelength-division multiplexing (WDM)
networks.  In recent years, EONs have been intensely researched by
both the industry and the academia.

% On slices and why they are better than wavelengths.

In EONs, the optical spectrum (the erbium window) is divided into thin
spectrum \emph{slices} (of, e.g., 6.25 GHz width), as opposed to
coarse fixed-grid channels (of, e.g., 25 GHz width) of
wavelength-division multiplexing (WDM) networks.  In EONs,
\emph{contiguous} slices are concatenated to form a \emph{slot}.
Slots are tailored for a specific demand, unlike WDM channels, thus
making EONs more spectrum-efficient than WDM networks.

% Routing of a single connection.

One of the many research problems of the EON design, planning, and
operation is \emph{the routing of a single connection in an EON},
which is the single most important operation of a network management
system (NMS).  For the NMS based on the Generalized Multi-Protocol
Label Switching (GMPLS), the path computation element (PCE) is
responsible for solving the routing and spectrum assignment (RSA)
problem for the given demand, i.e., finding the path and the slices.

% Solve RSA fast and well.

EONs should deal with dynamic traffic, where connections frequently
arrive and do not last long as opposed to the incumbent WDM traffic.
Furthermore, given the growing optical networks, the ever-increasing
need for bandwidth and connection agility, further increased by the
requirements of the fifth generation (5G) wireless networks, the PCE
is expected to solve the RSA problem fast and well.

% On constriction.

The RSA problem can be constrained by, for example, the path length,
i.e., we can require the solution to be shorter than the given limit.
This is an acceptable and desired limitation, since we may need to
limit the path length for a number of reasons: the connection takes
too much of network resources, the latency is too large, or the
quality of the optical signal is low.  The constriction limits the
search space, thus perhaps making the problem tractable, i.e.,
solvable in reasonable time.  To the best of our knowledge, it has not
been proven whether a constriction makes the RSA problem tractable or
not.

% Contribution.

Our novel contribution is the algorithm which quickly and optimally
solves the constrained RSA problem for a single demand.  The algorithm
is an adaptation and constriction of the Dijkstra shortest path
algorithm.  We show its effectiveness in comparison to routing with
the edge-disjoint shortest paths and the Yen $K$ shortest paths.  The
high-quality, high-performance implementation of the algorithm using
the Boost Graph Library (BGL) is available at \cite{abcdwebsite} under
the General Public License (GPL).

% About Dijkstra

Dijkstra is a principal graph algorithm, amenable to various
adaptations due to its simple and clever design.  Dijkstra is
efficient and optimal, and follows the label-setting paradigm, as
opposed to the label-correcting paradigm \cite{networkflows}.  At
first look, our adaptation seems to divorce the label-setting paradigm
in favor of the label-correcting paradigm, because we allow for
revisiting nodes, which Dijkstra does not do, and which is a hallmark
of the label-correcting algorithms.  But this is not so, the proposed
algorithm is still a label-setting algorithm.

% Article organization.

The article is organized as follows.  In Section \ref{related} we
briefly review related works, in Section \ref{statement} we define the
research problem, in Section \ref{algorithm} we describe the
algorithm, and in Section \ref{simulations} we report on the
simulation results.  Finally, Section \ref{conclusion} concludes the
article.

%%%%%%%%%%%%%%%%%%%%%%%%%%%%%%%%%%%%%%%%%%%%%%%%%%%%%%%%%%%%%%%%%%%%%%%%%%%

\section{Related works}
\label{related}

The RSA problem is reported to be NP-complete, so along with linear
programming formulations, there have been heuristic algorithms
proposed for real-sized networks \cite{10.1109/JLT.2014.2315041}.  In
\cite{10.1109/SURV.2012.010912.00123}, these algorithms are
categorized into one-stage algorithms, which route and assign spectrum
in one stage, and two-stage algorithms, which do it in two separate
stages.

In \cite{10.1109/JLT.2014.2315041}, the authors propose a one-stage
heuristic algorithm, which is a constrained Yen $K$ shortest path
algorithm.  The algorithm prunes the path deviations incapable of
supporting a demand.  The algorithm resorts to the Dijkstra algorithm
to compute shortest paths.

In \cite{10.1364/NFOEC.2011.JWA055}, the authors report a two-stage
algorithm for routing with Yen $K$ shortest paths.  First, the Yen
algorithm computes $K$ shortest paths also using the Dijkstra
algorithm, and next they try to establish a connection along these
paths. The authors also proposed a one-stage algorithm, termed a
modified Dijkstra algorithm, which is a typical constriction of the
Dijkstra algorithm, where a candidate path is rejected if it cannot
support a demand due to the lack of slices.

Another two-stage algorithm is routing with edge-disjoint shortest
paths.  First, all edge-disjoint shortest paths are found with the
Dijkstra algorithm, and next they try to establish a connection along
these paths.

All these algorithms, unlike ours, fail to find a shortest path
capable of supporting a demand, when there is a shorter path incapable
of supporting a demand, because that shorter path decoys Dijkstra into
a dead end.

%%%%%%%%%%%%%%%%%%%%%%%%%%%%%%%%%%%%%%%%%%%%%%%%%%%%%%%%%%%%%%%%%%%%%%%%%%%

\section{Problem statement}
\label{statement}

Given:

\begin{itemize}

\item directed multigraph $G = (V, E)$, where $V = \{v_i\}$ is a set
  of nodes, and $E = \{e_i\}$ is a set of edges,

\item attribute $c$ of edge $e_i$, i.e., $e_i.c$, which gives
  non-negative cost (length) of edge $e_i$,

\item attribute $SSC$ of edge $e_i$, i.e., $e_i.SSC$, which gives the
  set of available slices, with $\Omega$ being the set of all slices,

\item maximal path cost (length) $m$,

\item demand $d = (s, t, n)$, where $s$ is the source node, $t$ is
  the target node, and $n$ is the number of contiguous slices
  required.

\end{itemize}

Sought:

\begin{itemize}

\item shortest path $p = (e_1, ..., e_i, ..., e_l)$ for demand $d$ in
  graph $G$, where $e_i$ is the $i$-th edge of path $p$,

\item largest set of slices $\Sigma$, which can support demand $d$.
  
\end{itemize}

The objective is to find the \emph{largest} set of slices (SSC) along
the shortest path (SP).  An SSC can describe any slices available on
an edge or along a path, contiguous or not.  The sought SSC is the
largest possible, which can support demand $d$ with $n$ slices, i.e.,
it can have any number of contiguous spectrum fragments, each with at
least $n$ slices.

This problem formulation allows the algorithm to be more generic and
agnostic of the \emph{spectrum allocation policy}, because once the
largest SSC is found, any policy can be used to allocate slices, e.g.,
first or fittest.

%%%%%%%%%%%%%%%%%%%%%%%%%%%%%%%%%%%%%%%%%%%%%%%%%%%%%%%%%%%%%%%%%%%%%%%%%%%

\section{Proposed algorithm}
\label{algorithm}

% On the algortihm changes in general: adaptation and constriction.

We adapted and constrained the shortest path Dijkstra algorithm to
find an SP in EONs.  The adaptation is novel, and the constriction is
trivial.  The adaptation keeps track of the SSCs along the found
paths, while the constriction limits the length of an SP.  The found
paths are the SPs capable of supporting a given SSC, though they are
usually not the SPs in the graph.

% Dijkstra: label-setting, edge relaxation and label.

The Dijkstra algorithm is a \emph{label-setting} algorithm in that
once a node is visited, its label is set, and does not change, but the
label is updated by the \emph{edge relaxation}, if the given edge
yields a better than known label.  In label-setting algorithms, a
label is associated with every node, and gives information on what
cost and how to reach the given node from the source.  In Dijkstra a
label is the pair of cost and a preceding node.

% Definition of our label.

Our label, however, is more elaborate, since it has to describe more
elaborate data.  We define a label as \emph{a tuple of cost, a
preceding edge, and an SSC.}  For instance, label
$\nlabel{1}{e_1}{\SSC{1, 2}}$ says that a node is reached with cost 1
along edge $e_1$ and with the SSC of $\SSC{1, 2}$.  To allow for
multigraphs, in the tuple we keep a preceding edge, not a node.

% How Dijkstra compares labels.

In Dijkstra, node labels converge to their optimum by edge relaxation,
which updates a label when a better one if found.  Dijkstra compares
two labels: a candidate one, and a known one.  The candidate label is
better if it offers to reach the given node at a lower cost than the
known one.

% How we compare labels.

When trying to relax a candidate edge, we also compare two labels, but
we take into account not only the costs, but also the SSCs of the
labels.  Label $l_1$ is better than or equal to label $l_2$, denoted
by $l_1 \le l_2$, if $\text{cost}(l_1) \le \text{cost}(l_2)$ and
$SSC(l_1) \supseteq SSC(l_2)$.

% On our node labels.

In Dijkstra, a node has a single label, while we allow a node to have
a set of labels, provided that no label is better than or equal to
some other label, i.e., for any labels $l_i$ and $l_j$ of a given
node, $l_i \le l_j$ is false.  Our edge relaxation takes care of that.

% Path constriction.

As to the constriction of the path length, during a node visit,
Dijkstra traverses the out-edges of the node to find candidate labels,
and we require a candidate label to be dropped if its cost exceeds the
limit $m$.

% Reason for limiting the search space.

We had to limit the path length, because we had to narrow the solution
search space.  Otherwise, the algorithm can, as in some cases we ran
into, keep going through a very large search space for days, and not
find a solution.

\subsection{Adaptation}

The adaptation takes into account the spectrum continuity and
contiguity constraints.  The following three observations shaped the
adaptation.

\subsubsection{Revisit nodes}

In Dijkstra, a node is visited once for a single label.  We, however,
allow for revisiting nodes for multiple labels, because one of them
yields, if possible, an SP capable of supporting a given demand.

\begin{figure}
  \begin{tikzpicture}[node distance = 3 cm]
    \node [node] (s) {$s$};
    \node [node, right of = s] (i) {$i$};
    \node [node, right of = i] (t) {$t$};
    \path (s) edge [bend left] node [circle, fill = white] {$e_1$}
                               node [above] {\elabel{1}{\SSC{1, 2}}} (i);
    \path (s) edge [bend right] node [circle, fill = white] {$e_2$}
                                node [below] {\elabel{2}{\SSC{2, 3}}} (i);
    \path (i) edge node [circle, fill = white] {$e_3$}
                   node [above] {\elabel{10}{\SSC{2, 3}}} (t);
  \end{tikzpicture}
  \caption{Example for node revisiting and looping.}
  \label{f:example1}
\end{figure}
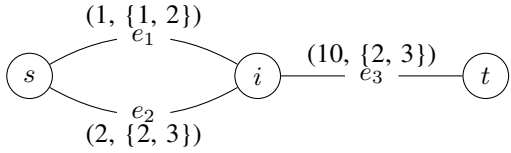

We show an example in Fig.~\ref{f:example1} to motivate node
revisiting, where the edge label gives a length and an SSC available
on an edge, e.g., \elabel{1}{\SSC{1, 2}} says the edge is of length 1
and slices 1 and 2 are available.  We are searching for an SP with two
slices from node $s$ to node $t$.

In the first step of Dijkstra, node $s$ is visited, and node $i$ is
discovered along the two parallel edges $e_1$ and $e_2$, but the label
is not updated for the longer edge $e_2$.  In the second step, node
$i$ is visited, and its label is set.  Now the final label for node
$i$ is known: $\nlabel{1}{e_1}{\SSC{1, 2}}$.  The problem is that node
$t$ cannot be discovered, because the spectrum continuity constraint
would be violated: the SSC of edge $e_3$ is $\SSC{2, 3}$, node $i$ was
reached with SSC $\SSC{1, 2}$, and the demand requires two slices.

\emph{Revisiting nodes solves this problem.  We allow for revisiting a
node even at a higher than known cost.}  In Dijkstra, in contrast, a
node is visited only once at the lowest cost.

Continuing with the example, and allowing for node revisiting, now
node $i$ is discovered along both parallel edges $e_1$ and $e_2$, and
none of the discoveries is discarded.  Then node $i$ is visited along
edge $e_1$ with label $\nlabel{1}{e_1}{\SSC{1, 2}}$, and then
revisited along edge $e_2$ with label $\nlabel{2}{e_2}{\SSC{2, 3}}$,
thus allowing node $t$ to be discovered, end eventually visited with
label $\nlabel{3}{e_3}{\SSC{2, 3}}$.

For simplicity, we illustrated node revisiting with parallel edges,
but could have also used parallel paths.  In the example, for
instance, edge $e_2$ can be replaced with two edges and a node
between them.

\subsubsection{Avoid loops}

Revisiting nodes may cause the search to find paths with loops.  For
instance, considering the same example in Fig.~\ref{f:example1}, when
visiting node $i$, we discover node $s$ and later revisit it, thus
finding the loop $(e_1, e_2)$.  In Dijkstra, loops are avoided by the
edge relaxation, which accepts only labels of lower cost: since edge
weights are non-negative, loops cannot decrease cost, and so they will
not be allowed by edge relaxation.  The thing is, that we need to
allow for revisiting even at higher costs, but also need to avoid
loops.

\emph{To avoid loops, and still to allow for node revisiting, an edge
can be relaxed even at a cost higher than the cost of any node
label, provided the candidate label offers an SSC not already
included in SSCs of the node labels.}  Therefore, a node is visited
and possibly revisited always at the lowest cost for an SSC not
included in the SSCs of previous visits.  And so, a node can have a
set of labels, but no label is better than or equal to some other
label, i.e., for any labels $l_i$ and $l_j$ of a given node, $l_i \le
l_j$ is false.

For example, in Fig.~\ref{f:example1}, the initial label for node $s$
is $l_1 = \nlabel{0}{\nulledge}{\FSSC}$.  The null edge $\nulledge$,
which is not present in graph $G$, marks the beginning of an SP.  When
visiting node $i$ with label $\nlabel{1}{e_1}{\SSC{1, 2}}$, node $s$
is discovered along edge $e_2$ with label $l_2 =
\nlabel{3}{e_2}{\SSC{2}}$, but the edge will not be relaxed, because
$l_1 \le l_2$, thus avoiding a loop.

\subsubsection{Purge labels}

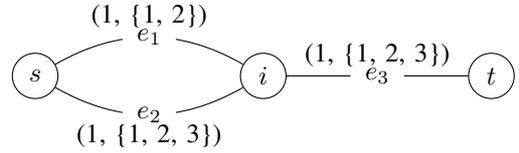
\begin{figure}
  \begin{tikzpicture}[node distance = 3 cm]
    \node [node] (s) {$s$};
    \node [node, right of = s] (i) {$i$};
    \node [node, right of = i] (t) {$t$};
    \path (s) edge [bend left] node [circle, fill = white] {$e_1$}
                               node [above] {\elabel{1}{\SSC{1, 2}}} (i);
    \path (s) edge [bend right] node [circle, fill = white] {$e_2$}
                                node [below] {\elabel{1}{\SSC{1, 2, 3}}} (i);
    \path (i) edge node [circle, fill = white] {$e_3$}
                   node [above] {\elabel{1}{\SSC{1, 2, 3}}} (t);
  \end{tikzpicture}
  \caption{Example for purging worse labels.}
  \label{f:example2}
\end{figure}

When we relax an edge, we add a new label for the node, but we may
also need to purge worse labels.  The purging of worse labels is
illustrated by the example in Fig.~\ref{f:example2}.  When visiting
node $s$, node $i$ is discovered along edge $e_1$ with label $l_1 =
\nlabel{1}{e_1}{\SSC{1, 2}}$, and node $i$ has label $l_1$ only.
Next, node $i$ is discovered along edge $e_2$ with label $l_2 =
\nlabel{1}{e_2}{\SSC{1, 2, 3}}$.  Label $l_2$ is better than $l_1$,
because the SSC of $l_2$ includes the SSC of $l_1$, and both labels
are of the same cost.  We purge $l_1$ from the set of labels of node
$i$, and now node $i$ has label $l_2$ only.

\subsection{Algorithm}

Algorithm \ref{a:algorithm} presents the complete algorithm with the
typical of the Dijkstra algorithm structure, where the main loop
processes the priority queue elements.  Our priority queue $Q$ stores
the elements $q = (c, e)$, which are pairs of cost $c$, and edge $e$.
The queue is sorted according only to the increasing cost $c$ of the
elements, without the consideration of edge $e$, which is associated
with the cost to tell what to process.

We require $Q$ to store only unique elements: pushing the same element
$q$ many times results in just one element $q$ in the queue.  This
property is required to process in one iteration of the main loop all
the labels with the same cost and edge, but a different SSC.  $Q$ can
be implemented as a set to guarantee this property.

We initialize $L_s = \{\nlabel{0}{\nulledge}{\Omega}\}$ to make all
slices available at node $s$ at cost 0.  The null edge $\nulledge$,
which is not present in graph $G$, marks the beginning of an SP.
$L_v$ is a set of labels of node $v$, and $L = \{L_v\}$ is the set of
sets of node labels.  Next, we put element $\elabel{0}{e_\emptyset}$
to the queue to boot the search.

In every iteration of the main loop, we pop from $Q$ element $q = (c,
e)$, and visit node $v = e.target$ reached along edge $e$ at cost $c$.
The target node of an edge is given by attribute $target$, i.e.,
$e.target$, with the special case of $e_{\emptyset}.target = s$.  If
$v = t$, then we found a solution and break the main loop.

In the main loop we iterate over two nested loops.  One loop iterates
over all SSCs $S$ of labels in $L_v$ with cost $c$ and edge $e$, and
the other loop iterates over all outgoing edges $e'$ of node $v$, in
order to discover a neighbor node $v' = e'.target$ along edge $e'$ at
cost $c' = c + e'.cost$ with SSC $S' = S \cap e'.SSC$.  We continue
working with $S'$ and $e'$, if $c' \le m$ and $S'$ can support demand
$d$, i.e., $S'$ has at least $n$ contiguous slices.

Next, we check whether edge $e'$ can be relaxed with candidate label
$l'$, i.e., whether node $v'$ has no label $l$ better than or equal to
label $l'$.  If so, then
\begin{inparaenum}
\item we purge every label $l$ of node $v'$ if $l' \le l$,
\item add label $l'$ to the set of labels of node $v'$,
\item push element $(c', e')$ to Q.
\end{inparaenum}
Edge relaxation replenishes the queue, and the algorithm keeps
iterating until destination node $t$ is reached, or the queue is
empty.

Finally, function $trace$ traces back an SP found, if any, based on
the node labels $L$.  We do not present the algorithm for tracing back
an SP, since it is rather easy.

\begin{algorithm}[t]
  \caption{\\
    In: $G = (V = \{v_i\}, E = \{e_i\}), W(e_i), S(e_i), l, d = (s, t, n)$\\
    Out: $p = (e_1, ..., e_i, ..., e_l)$, $\Sigma = \{\sigma_i\}$}
  \label{a:algorithm}
    \begin{algorithmic}
      \STATE $L_s = \{(0, \nulledge, \Omega)\}$
      \STATE push $(0, e_\emptyset)$ to $Q$ 
      \WHILE{$Q$ is not empty}
      \STATE $q = (c, e) = \textrm{pop}(Q)$
      \STATE $v = e.target$
      \IF{$v == t$}
      \STATE break the while loop
      \ENDIF
      \STATE $SSSC = \{l.SSC: l \in L_v \text{ and } l.c == c \text{ and } l.e == e\}$
      \FORALL{$S \in SSSC$}
      \FORALL{$e' \in \text{outgoing edges of }v$}
      \STATE $S' = S \cap S(e')$
      \STATE $c' = c + W(e')$
      \IF{$c' \le m \text{ and } S' \text{ can support $d$}$}
      \STATE $v' = e'.target$
      \STATE $l' = (c', e', S')$
      \IF{$\nexists l \in L_{v'} : l \le l'$}
      \STATE $L_{v'} = L_{v'} \setminus \{l: l \in L_{v'} \text{ and } l' \le l\}$
      \STATE $L_{v'} = L_{v'} \cup \{l'\}$
      \STATE push $(c', e')$ to $Q$
      \ENDIF
      \ENDIF
      \ENDFOR
      \ENDFOR
      \ENDWHILE
      \RETURN $(p, \Sigma) = trace(L, s, t)$
    \end{algorithmic}
\end{algorithm}

%%%%%%%%%%%%%%%%%%%%%%%%%%%%%%%%%%%%%%%%%%%%%%%%%%%%%%%%%%%%%%%%%%%%%%%%%%%

\section{Simulative studies}
\label{simulations}

% What's the deal with the simulations.

We evaluate the performance of the proposed algorithm with
simulations, and compare it to the performance of routing with the
edge-disjoint paths and the Yen $K$ shortest paths.  We also show,
based on the simulation results, that the proposed algorithm
efficiently solves the constrained RSA problem, which suggests that
this problem is tractable, though we offer no proof.

% The Edge-disjoint paths and the Yen K shortest paths.

For comparison, we use the edge-disjoint paths and the Yen $K$
shortest paths, because they are very different: the edge-disjoint
shortest paths do not share even a single edge, while the Yen $K$
shortest paths can differ with a single edge only.  To find
edge-disjoint paths, we search for an SP in a graph with the edges of
the previous SPs disabled.  The Yen $K$ shortest paths are found with
the well-known Yen algorithm.  We do not limit the number of
edge-disjoint paths, because at most it equals to the degree of the
source or target nodes, which is a small number.  We limit, however,
the Yen shortest paths to at most $K = 10$, because Yen can produce a
very large number of paths.

% Routing with the paths.

Having either the edge-disjoint shortest paths or the Yen K shortest
paths, we try to route a demand as follows.  We start with the first
SP, and calculate the largest available SSC along the path by
intersecting all the SSCs available on every edge of the shortest
path.  If the largest SSC cannot support the demand, we try the next
shortest path, until there are no more shortest paths.

% Spectrum allocation policy.

Having found a path with the largest SSC, a spectrum allocation policy
allocates $n$ slices for demand $d$ from the largest SSC.  We consider
only the \emph{fittest} and the \emph{first} spectrum allocation
policies.  The fittest policy allocates $n$ slices in the fittest
fragment of the largest SSC, which can support demand $d$, i.e., the
smallest fragment with at least $n$ slices.  The first policy
allocates $n$ slices in the first fragment (with slices of the
smallest numbers) of the largest SSC, which can support demand $d$.

To be general, we formulate and solve the problem for a directed
multigraph, but we use it for an undirected graph, which models an
EON.

\subsection{Simulation setting}
\label{setting}

% On random graphs and traffic.

We generate a set of random graphs with random traffic to obtain
reliable statistical results for various populations of interest,
because we find studying a specific topology (e.g., Polish PIONIER or
NSFNet) with some specific traffic case rather inconclusive.  We use
Gabriel graphs, because they have been shown to model the properties
(e.g., the node degree) of the transport networks very well
\cite{10.1109/ICUMT.2013.6798402}.

% On the Gabriel graphs we generate.

We randomly generate 50 Gabriel graphs, where each edge has 400
slices.  Each graph has 100 nodes, which are uniformly distributed
over an area 1000 km long and 1000 km wide.  In generating Gabriel
graphs, the number of edges cannot be directly controlled, as it
depends on the location of nodes, and on the candidate edges meeting
the conditions of the Gabriel graph.  The statistics of the generated
graphs are given in Table \ref{t:netstats}.  The limit on the path
length is $m = 2000$ km, which is well above 1582 km, the length of
the longest of all shortest paths in the generated graphs.

% On traffic in general.

Demands arrive according to the exponential distribution with the rate
of $\lambda$ demands per day.  The probability distribution of the
demand holding time is also exponential with the mean of $\beta = 10$
hours.  The number of slices a demand requires follows the Poisson
distribution with the mean of 10 slices.

% Traffic is irrelevant.  Don't need defragmentation.

We argue that the choice of a traffic model is irrelevant to our study
as the traffic only produces the input data (i.e., the state of the
graph) for the routing algorithms, and we chose the exponential and
Poisson distributions to keep the discussion simple.  The question is
how the algorithms perform under the given utilization and
fragmentation, regardless of how the utilization and fragmentation
were obtained, which could have been equally well produced randomly.
For the very same reason we do not incorporate into our study spectrum
defragmentation.

% On network utilization.

We define \emph{network utilization} as the ratio of the number of the
slices in use to the total number of slices on all edges.  We cannot
directly control the network utilization, but measure it in response
to the offered load.  And so to obtain different values of network
utilization, we varied $\lambda$ with 27 different values of 10, 12.5,
15, 17.5, 20, 25, 30, 35, 40, 45, 50, 55, 60, 70, 80, 90, 100, 150,
200, 300, 400, 500, 600, 700, 800, 900, and 1000 demand arrivals per
day.

% What a single simulation did.

\emph{A simulation run} simulates 100 days of a network in operation.
For every simulated day, the following values are measured:
\begin{inparaenum}[(a)]
\item the instantaneous network utilization,
\item the average probability of establishing a connection during that day,
\item the instantaneous number of active connections,
\item the instantaneous amount of capacity served,
\item the average length of an established connection during that day,
\item the average number of slices of an established connection during that day,
\item the instantaneous number of spectrum fragments of an edge,
\item the average time taken by an SP search during that day.
\end{inparaenum}
When a simulation finishes, the measured values are averaged and
reported as the simulation results.

% The population results.

We want to get reliable results for a statistical population of
simulation runs.  A population is described by the routing algorithm
used, the spectrum allocation policy used, and $\lambda$, and so there
are 162 populations considered (three routing algorithms $\times$ two
spectrum allocation policies $\times$ 27 values of $\lambda$).  In a
given population, all simulation runs have the same parameters, except
the seed of a random number generator in order to generate different
Gabriel graphs and different traffic.  To get reliable results for a
population, we carry out 50 simulation runs which are the population
samples, and calculate the sample means of all the results reported by
a simulation, except the average search time of which we take the
minimum.  In total there are 8100 simulation runs (162 populations
$\times$ 50 samples).  We reckon the sample means reliably estimate
the results of their populations, since their relative standard error
is below 1\%.

% On the average search time.

We do not calculate the sample mean of the average search time, but
take the minimum, since we run simulations using a supercomputing
infrastructure, and cannot control the specific hardware for our
simulations, and how much the hardware is loaded with other jobs.
Other processes running can heavily utilize memory, thus causing cache
misses in our simulations, which severely degrade performance.  The
supercomputing infrastructure is composed of thousands of nodes
equipped with the state-of-the-art multi-core processors of the AMD64
architecture.

\begin{table}
  \caption{Statistics of the generated Gabriel networks.}
  \label{t:netstats}
  \begin{tabular}{|l|r|r|r|r|}
    \hline
    {\bf{}value} & {\bf{}min} & {\bf{}average} & {\bf{}max} & {\bf{}variance}\\
    \hline
    Number of links & 160 & 179.2 & 194 & 48.52\\
    Link length & 1 & 97.95 & 347 & 2696.46\\
    Node degree & 1 & 3.584 & 8 & 1.2213\\
    SP length & 1 & 589.61 & 1582 & 78118.1\\
    SP hops min & 1 & 6.7634 & 22 & 10.8737\\
    \hline
  \end{tabular}
\end{table}

%%%%%%%%%%%%%%%%%%%%%%%%%%%%%%%%%%%%%%%%%%%%%%%%%%%%%%%%%%%%%%%%%%%%%%%%%%%

\subsection{Simulation results}
\label{results}

\begin{figure*}
  \begin{tabular}{rp{0.5 cm}r}
    \subfloat[The probability of establishing a connection.]{%
      \label{f:pec}%
      \begin{tikzpicture}
\begin{axis}[xlabel = network utilization, ylabel style = {align = center}, ylabel = {probability}, legend columns = 3, legend to name = legend,
height = 5.25 cm, width = 8 cm,
scaled y ticks = false,
]
\addplot[solid, very thick, red, mark size = 1 pt, mark options = {solid, thin}]
coordinates {
(0.086705629702, 1.0) (0.108321141216, 1.0) (0.12991051779, 0.9999789472) (0.151810419134, 0.9999046442) (0.17304654392, 0.999603033) (0.21673848926, 0.9970970046) (0.25923923948, 0.9887169924) (0.29774809288, 0.9693344378) (0.32696349078, 0.9426652376) (0.35070618776, 0.9090997458) (0.36838103266, 0.8746160524) (0.38137073416, 0.843152826) (0.39268851456, 0.8132496084) (0.41026349892, 0.758198448) (0.4238232238, 0.7133272532) (0.43561990178, 0.673761892) (0.44446794196, 0.6402445092) (0.47833246, 0.5176793236) (0.4999098002, 0.4445479644) (0.5295011246, 0.3566588778) (0.550188323, 0.3038360008) (0.5662778028, 0.2692026552) (0.5791035826, 0.2430689584) (0.5903855586, 0.2234743116) (0.5997335402, 0.2071849668) (0.608103685, 0.1938133052) (0.6159557148, 0.1828414832) 
};
\addlegendentry{proposed, fittest}
\addplot[dashed, very thick, red, mark size = 1 pt, mark options = {solid, thin}]
coordinates {
(0.086663621422, 0.9998491442) (0.107328993256, 0.9991253184) (0.12876481919, 0.9962865052) (0.148775796434, 0.9911094764) (0.1675690556, 0.9808155724) (0.19810646238, 0.952251716) (0.22270764192, 0.9174637686) (0.24198303742, 0.8802881902) (0.2581683898, 0.8423269368) (0.27085223244, 0.8087467468) (0.28255796086, 0.7777205092) (0.29361976816, 0.749895367) (0.30168320924, 0.7245583388) (0.31645890366, 0.6808492194) (0.32990244492, 0.6415935874) (0.34151160912, 0.609981759) (0.35131420306, 0.5801670338) (0.3877354178, 0.4808692718) (0.4143904902, 0.4182004168) (0.4513921192, 0.341061997) (0.4788833498, 0.2943961766) (0.5000962058, 0.26265875) (0.5173826246, 0.2387836138) (0.5322993904, 0.2201314074) (0.5455395396, 0.2049567288) (0.557242117, 0.1926717702) (0.5673326562, 0.1817769586) 
};
\addlegendentry{edksp, fittest}
\addplot[dotted, very thick, red, mark size = 1 pt, mark options = {solid, thin}]
coordinates {
(0.086355280082, 0.9998179294) (0.107100593176, 0.9984475148) (0.12728235469, 0.994428539) (0.147537033634, 0.9862346908) (0.1629804652, 0.9754836176) (0.19373712338, 0.9437023558) (0.21668548612, 0.912019801) (0.23710054442, 0.8779552032) (0.252270374, 0.8470574992) (0.26813430504, 0.8164687356) (0.28014051126, 0.7897361774) (0.29206624636, 0.7611208634) (0.30137254424, 0.7378027408) (0.31943920006, 0.695975261) (0.33380474712, 0.6586636452) (0.34630187152, 0.6270060886) (0.35840043126, 0.5991084792) (0.4002842278, 0.4966621844) (0.429382547, 0.433302999) (0.4706603972, 0.3542115266) (0.5006188356, 0.3051437308) (0.5231465794, 0.2714204674) (0.5414597174, 0.2466979542) (0.5572275366, 0.2267090244) (0.5703908678, 0.2107550598) (0.5823302018, 0.1974180354) (0.5919572576, 0.1865258966) 
};
\addlegendentry{yenksp, fittest}
\addplot[solid, mark size = 1 pt, mark options = {solid, thin}]
coordinates {
(0.086702837542, 1.0) (0.108298397536, 1.0) (0.12994017697, 0.999943492) (0.151802982934, 0.999968117) (0.17408250252, 0.9995949038) (0.21837125666, 0.9960545142) (0.26142735368, 0.9866256488) (0.29739790528, 0.968934748) (0.32663134678, 0.9401832238) (0.34507561722, 0.9075594362) (0.36621243026, 0.8748287366) (0.37911476176, 0.8408267414) (0.39162461916, 0.8096912374) (0.40909354352, 0.7561908122) (0.4220238248, 0.7108790594) (0.43338709998, 0.669385225) (0.44216048956, 0.6372082816) (0.475437044, 0.5165586634) (0.4966150342, 0.442690156) (0.5265456358, 0.355161418) (0.5467033538, 0.303048446) (0.5619558026, 0.267604421) (0.5754145046, 0.2415686062) (0.5862118814, 0.2220419158) (0.5958195578, 0.2059258082) (0.60401357, 0.1925964774) (0.6114761976, 0.1814615234) 
};
\addlegendentry{proposed, first}
\addplot[dashed, mark size = 1 pt, mark options = {solid, thin}]
coordinates {
(0.086665838862, 0.9998949184) (0.107918635556, 0.999094187) (0.12954746983, 0.9972662052) (0.149540714834, 0.9920522298) (0.168034899, 0.983129536) (0.20096933738, 0.9532778608) (0.22513554252, 0.9168745476) (0.24398380202, 0.880994329) (0.2600443686, 0.8436296618) (0.27252377724, 0.8107109058) (0.28406668726, 0.7795337196) (0.29454112636, 0.7497768854) (0.30349057124, 0.722725504) (0.31773460946, 0.6797443732) (0.33089523912, 0.6402682772) (0.34164904432, 0.608794238) (0.35105451726, 0.58120897) (0.3878739872, 0.4798169256) (0.413792956, 0.417013415) (0.4509418896, 0.3403783692) (0.4774484556, 0.2933198532) (0.4978965204, 0.2615230512) (0.5152316676, 0.2378619408) (0.530098128, 0.2188741556) (0.5432291, 0.203590597) (0.5539268922, 0.1913582238) (0.564077479, 0.180817159) 
};
\addlegendentry{edksp, first}
\addplot[dotted, mark size = 1 pt, mark options = {solid, thin}]
coordinates {
(0.086515602382, 0.9997901514) (0.107834904476, 0.9980029504) (0.12740081217, 0.9950303706) (0.146206386034, 0.9863345942) (0.1639753978, 0.9743326666) (0.19324621898, 0.9440636072) (0.21717579772, 0.9123624126) (0.23627081222, 0.8778093596) (0.253481025, 0.84483387) (0.26743315404, 0.8150018808) (0.28020710606, 0.786143067) (0.29107584596, 0.7594759206) (0.30031955864, 0.7356997006) (0.31905951866, 0.692486026) (0.33359707252, 0.6571344456) (0.34584809692, 0.6259865008) (0.35663481766, 0.5963644706) (0.3985821714, 0.4957503226) (0.4276507036, 0.4318444822) (0.46878804, 0.352585532) (0.4973622534, 0.3041306542) (0.5201774492, 0.2700937736) (0.5381127192, 0.2449492688) (0.5530367688, 0.2253466716) (0.5654528078, 0.2096881168) (0.577790159, 0.196417257) (0.5875427202, 0.1853562596) 
};
\addlegendentry{yenksp, first}
\end{axis}
\end{tikzpicture}}&&%
    \subfloat[The time of shortest path search.]{%
      \label{f:spsat}%
      \begin{tikzpicture}
\begin{axis}[xlabel = network utilization, ylabel style = {align = center}, ylabel = {time [s]}, legend columns = 3, legend to name = legend,
height = 5.25 cm, width = 8 cm,
scaled y ticks = false,
]
\addplot[solid, very thick, red, mark size = 1 pt, mark options = {solid, thin}]
coordinates {
(0.086705629702, 0.37658156) (0.108321141216, 0.353297369) (0.12991051779, 0.35752177) (0.151810419134, 0.259658551) (0.17304654392, 0.233553397) (0.21673848926, 0.159488399) (0.25923923948, 0.124097649) (0.29774809288, 0.083918497) (0.32696349078, 0.083786677) (0.35070618776, 0.064026713) (0.36838103266, 0.069223613) (0.38137073416, 0.054416067) (0.39268851456, 0.047666983) (0.41026349892, 0.052378154) (0.4238232238, 0.044795233) (0.43561990178, 0.03999202) (0.44446794196, 0.039262262) (0.47833246, 0.027502958) (0.4999098002, 0.021098398) (0.5295011246, 0.0155497651) (0.550188323, 0.0127173644) (0.5662778028, 0.0101319303) (0.5791035826, 0.0087320842) (0.5903855586, 0.0080193328) (0.5997335402, 0.0072192592) (0.608103685, 0.0065686862) (0.6159557148, 0.0061634157) 
};
\addlegendentry{proposed, fittest}
\addplot[dashed, very thick, red, mark size = 1 pt, mark options = {solid, thin}]
coordinates {
(0.086663621422, 0.0020531323) (0.107328993256, 0.0019432428) (0.12876481919, 0.0019190695) (0.148775796434, 0.0019862336) (0.1675690556, 0.0019542941) (0.19810646238, 0.0018734272) (0.22270764192, 0.0019922348) (0.24198303742, 0.0019108865) (0.2581683898, 0.0019079146) (0.27085223244, 0.0018745261) (0.28255796086, 0.0018727781) (0.29361976816, 0.0018421538) (0.30168320924, 0.0018003321) (0.31645890366, 0.0018012052) (0.32990244492, 0.0018162902) (0.34151160912, 0.0017307058) (0.35131420306, 0.0018202905) (0.3877354178, 0.001774741) (0.4143904902, 0.0016976283) (0.4513921192, 0.0016419875) (0.4788833498, 0.0014666174) (0.5000962058, 0.0015244543) (0.5173826246, 0.0015169783) (0.5322993904, 0.0014474626) (0.5455395396, 0.0013463693) (0.557242117, 0.0013178057) (0.5673326562, 0.0013814241) 
};
\addlegendentry{edksp, fittest}
\addplot[dotted, very thick, red, mark size = 1 pt, mark options = {solid, thin}]
coordinates {
(0.086355280082, 0.0159519803) (0.107100593176, 0.0158027943) (0.12728235469, 0.01615003) (0.147537033634, 0.015687082) (0.1629804652, 0.015735272) (0.19373712338, 0.015833671) (0.21668548612, 0.015877213) (0.23710054442, 0.015781727) (0.252270374, 0.016308206) (0.26813430504, 0.016352666) (0.28014051126, 0.016476058) (0.29206624636, 0.016488065) (0.30137254424, 0.016497134) (0.31943920006, 0.016451977) (0.33380474712, 0.016464648) (0.34630187152, 0.016546967) (0.35840043126, 0.016219575) (0.4002842278, 0.016701066) (0.429382547, 0.016700671) (0.4706603972, 0.016819394) (0.5006188356, 0.016745696) (0.5231465794, 0.016744547) (0.5414597174, 0.016525954) (0.5572275366, 0.016603159) (0.5703908678, 0.01571968) (0.5823302018, 0.016397923) (0.5919572576, 0.01603204) 
};
\addlegendentry{yenksp, fittest}
\addplot[solid, mark size = 1 pt, mark options = {solid, thin}]
coordinates {
(0.086702837542, 0.38192325) (0.108298397536, 0.32407842) (0.12994017697, 0.29157747) (0.151802982934, 0.22119623) (0.17408250252, 0.224808273) (0.21837125666, 0.145671516) (0.26142735368, 0.10968886) (0.29739790528, 0.088375301) (0.32663134678, 0.076086146) (0.34507561722, 0.062569169) (0.36621243026, 0.067418211) (0.37911476176, 0.050960633) (0.39162461916, 0.047981959) (0.40909354352, 0.04763558) (0.4220238248, 0.038769635) (0.43338709998, 0.038671628) (0.44216048956, 0.035425622) (0.475437044, 0.026131573) (0.4966150342, 0.021445489) (0.5265456358, 0.0152292941) (0.5467033538, 0.0120728488) (0.5619558026, 0.0101304229) (0.5754145046, 0.0087261125) (0.5862118814, 0.0079932197) (0.5958195578, 0.0072141058) (0.60401357, 0.006166846) (0.6114761976, 0.0061031744) 
};
\addlegendentry{proposed, first}
\addplot[dashed, mark size = 1 pt, mark options = {solid, thin}]
coordinates {
(0.086665838862, 0.002059301) (0.107918635556, 0.0019197933) (0.12954746983, 0.0019932652) (0.149540714834, 0.001901656) (0.168034899, 0.0018999689) (0.20096933738, 0.0019082527) (0.22513554252, 0.0019126163) (0.24398380202, 0.0018931373) (0.2600443686, 0.0019043154) (0.27252377724, 0.0018201785) (0.28406668726, 0.0018171444) (0.29454112636, 0.0018379403) (0.30349057124, 0.0019319704) (0.31773460946, 0.0018295333) (0.33089523912, 0.0018066112) (0.34164904432, 0.0018260483) (0.35105451726, 0.0017571567) (0.3878739872, 0.0017544434) (0.413792956, 0.0016419281) (0.4509418896, 0.001632009) (0.4774484556, 0.001523947) (0.4978965204, 0.0016165691) (0.5152316676, 0.0015434397) (0.530098128, 0.0015392413) (0.5432291, 0.0014567365) (0.5539268922, 0.0014201539) (0.564077479, 0.0013983006) 
};
\addlegendentry{edksp, first}
\addplot[dotted, mark size = 1 pt, mark options = {solid, thin}]
coordinates {
(0.086515602382, 0.015797764) (0.107834904476, 0.015825274) (0.12740081217, 0.015877224) (0.146206386034, 0.015842165) (0.1639753978, 0.016016401) (0.19324621898, 0.015345711) (0.21717579772, 0.016116508) (0.23627081222, 0.016218705) (0.253481025, 0.016511225) (0.26743315404, 0.016367481) (0.28020710606, 0.01638371) (0.29107584596, 0.016519411) (0.30031955864, 0.016603873) (0.31905951866, 0.016429218) (0.33359707252, 0.016672102) (0.34584809692, 0.017240887) (0.35663481766, 0.016814197) (0.3985821714, 0.01634544) (0.4276507036, 0.016782723) (0.46878804, 0.016537755) (0.4973622534, 0.016418946) (0.5201774492, 0.016627363) (0.5381127192, 0.016318786) (0.5530367688, 0.016477406) (0.5654528078, 0.015921005) (0.577790159, 0.016428501) (0.5875427202, 0.016275233) 
};
\addlegendentry{yenksp, first}
\end{axis}
\end{tikzpicture}}\\[20pt]
    \subfloat[The length of an established connection.]{%
      \label{f:lenec}%
      \begin{tikzpicture}
\begin{axis}[xlabel = network utilization, ylabel style = {align = center}, ylabel = {length [km]}, legend columns = 3, legend to name = legend,
height = 5.25 cm, width = 8 cm,
scaled y ticks = false,
]
\addplot[solid, very thick, red, mark size = 1 pt, mark options = {solid, thin}]
coordinates {
(0.086705629702, 595.609535) (0.108321141216, 596.634798) (0.12991051779, 597.2686568) (0.151810419134, 599.988476) (0.17304654392, 604.1250862) (0.21673848926, 615.9068358) (0.25923923948, 632.6454532) (0.29774809288, 649.5395916) (0.32696349078, 659.3440138) (0.35070618776, 663.7393572) (0.36838103266, 662.3014034) (0.38137073416, 659.5606932) (0.39268851456, 653.6678196) (0.41026349892, 642.7362248) (0.4238232238, 631.9673734) (0.43561990178, 620.8216376) (0.44446794196, 608.598706) (0.47833246, 560.248497) (0.4999098002, 524.4646122) (0.5295011246, 472.2659098) (0.550188323, 437.4627856) (0.5662778028, 410.6374668) (0.5791035826, 389.0835406) (0.5903855586, 372.3331474) (0.5997335402, 357.3035398) (0.608103685, 345.0859058) (0.6159557148, 333.7921444) 
};
\addlegendentry{proposed, fittest}
\addplot[dashed, very thick, red, mark size = 1 pt, mark options = {solid, thin}]
coordinates {
(0.086663621422, 595.5569778) (0.107328993256, 596.1680634) (0.12876481919, 599.1107738) (0.148775796434, 600.6685886) (0.1675690556, 601.2564808) (0.19810646238, 599.9964456) (0.22270764192, 596.5390134) (0.24198303742, 588.2927456) (0.2581683898, 578.2566006) (0.27085223244, 568.867818) (0.28255796086, 560.0888972) (0.29361976816, 550.433135) (0.30168320924, 542.2810716) (0.31645890366, 526.7579126) (0.32990244492, 512.3514196) (0.34151160912, 499.580091) (0.35131420306, 488.1723434) (0.3877354178, 445.0160876) (0.4143904902, 416.1090202) (0.4513921192, 375.8628338) (0.4788833498, 349.8926266) (0.5000962058, 330.109451) (0.5173826246, 315.0933748) (0.5322993904, 302.7229466) (0.5455395396, 292.0996968) (0.557242117, 283.2592658) (0.5673326562, 275.858917) 
};
\addlegendentry{edksp, fittest}
\addplot[dotted, very thick, red, mark size = 1 pt, mark options = {solid, thin}]
coordinates {
(0.086355280082, 595.2567884) (0.107100593176, 594.4358916) (0.12728235469, 593.5259562) (0.147537033634, 593.9071476) (0.1629804652, 589.8096016) (0.19373712338, 582.7115068) (0.21668548612, 574.4240414) (0.23710054442, 566.3284134) (0.252270374, 555.9787156) (0.26813430504, 547.5152182) (0.28014051126, 540.0543468) (0.29206624636, 530.4618086) (0.30137254424, 524.1889984) (0.31943920006, 510.3757128) (0.33380474712, 497.829806) (0.34630187152, 487.5363432) (0.35840043126, 477.7534834) (0.4002842278, 439.8851516) (0.429382547, 413.4201662) (0.4706603972, 377.3148466) (0.5006188356, 353.6439386) (0.5231465794, 335.5836468) (0.5414597174, 321.5680936) (0.5572275366, 310.7095068) (0.5703908678, 301.053732) (0.5823302018, 292.049402) (0.5919572576, 285.1685906) 
};
\addlegendentry{yenksp, fittest}
\addplot[solid, mark size = 1 pt, mark options = {solid, thin}]
coordinates {
(0.086702837542, 595.5954578) (0.108298397536, 596.5820646) (0.12994017697, 597.3832112) (0.151802982934, 600.3699358) (0.17408250252, 604.5130096) (0.21837125666, 617.2404658) (0.26142735368, 635.8557208) (0.29739790528, 650.1847988) (0.32663134678, 659.2487652) (0.34507561722, 661.4471566) (0.36621243026, 662.0332202) (0.37911476176, 658.6695598) (0.39162461916, 654.1970168) (0.40909354352, 641.6084808) (0.4220238248, 632.0027362) (0.43338709998, 619.3561684) (0.44216048956, 607.4301812) (0.475437044, 560.3194348) (0.4966150342, 524.0750798) (0.5265456358, 472.827576) (0.5467033538, 438.0116832) (0.5619558026, 411.8677378) (0.5754145046, 391.1602516) (0.5862118814, 374.877773) (0.5958195578, 359.8386322) (0.60401357, 347.9267944) (0.6114761976, 338.0729566) 
};
\addlegendentry{proposed, first}
\addplot[dashed, mark size = 1 pt, mark options = {solid, thin}]
coordinates {
(0.086665838862, 595.702175) (0.107918635556, 596.3999944) (0.12954746983, 598.6766418) (0.149540714834, 600.8483038) (0.168034899, 603.0505432) (0.20096933738, 603.9822352) (0.22513554252, 599.6870714) (0.24398380202, 590.3691006) (0.2600443686, 582.674268) (0.27252377724, 572.008709) (0.28406668726, 562.3687684) (0.29454112636, 554.1101204) (0.30349057124, 544.8401006) (0.31773460946, 529.6320176) (0.33089523912, 515.502984) (0.34164904432, 504.2164236) (0.35105451726, 493.7030968) (0.3878739872, 449.1929086) (0.413792956, 420.1608526) (0.4509418896, 380.3377122) (0.4774484556, 353.9240118) (0.4978965204, 334.2662842) (0.5152316676, 318.792422) (0.530098128, 306.3822928) (0.5432291, 296.2396458) (0.5539268922, 287.1827654) (0.564077479, 279.6132166) 
};
\addlegendentry{edksp, first}
\addplot[dotted, mark size = 1 pt, mark options = {solid, thin}]
coordinates {
(0.086515602382, 595.1750998) (0.107834904476, 594.6787334) (0.12740081217, 593.6607046) (0.146206386034, 592.4461988) (0.1639753978, 589.8577268) (0.19324621898, 583.877614) (0.21717579772, 575.7756336) (0.23627081222, 566.5765812) (0.253481025, 557.4316672) (0.26743315404, 548.8350142) (0.28020710606, 540.791727) (0.29107584596, 532.5644564) (0.30031955864, 524.71237) (0.31905951866, 511.6645672) (0.33359707252, 499.3203046) (0.34584809692, 489.1580934) (0.35663481766, 478.5718268) (0.3985821714, 441.6261364) (0.4276507036, 415.635423) (0.46878804, 379.4072212) (0.4973622534, 355.8461266) (0.5201774492, 337.6981566) (0.5381127192, 324.288163) (0.5530367688, 312.4831038) (0.5654528078, 303.2330924) (0.577790159, 294.741486) (0.5875427202, 287.7395232) 
};
\addlegendentry{yenksp, first}
\end{axis}
\end{tikzpicture}}&&%
    \subfloat[The number of slices of an established connection.]{%
      \label{f:nscec}%
      \begin{tikzpicture}
\begin{axis}[xlabel = network utilization, ylabel style = {align = center}, ylabel = {number of slices}, legend columns = 3, legend to name = legend,
height = 5.25 cm, width = 8 cm,
scaled y ticks = false,
]
\addplot[solid, very thick, red, mark size = 1 pt, mark options = {solid, thin}]
coordinates {
(0.086705629702, 10.015286926) (0.108321141216, 10.02090999) (0.12991051779, 10.018551406) (0.151810419134, 10.016939958) (0.17304654392, 10.013059856) (0.21673848926, 9.999975626) (0.25923923948, 9.960657434) (0.29774809288, 9.879996216) (0.32696349078, 9.78743758) (0.35070618776, 9.670684062) (0.36838103266, 9.569778704) (0.38137073416, 9.467368874) (0.39268851456, 9.399734152) (0.41026349892, 9.251608312) (0.4238232238, 9.146113096) (0.43561990178, 9.070708854) (0.44446794196, 9.00045412) (0.47833246, 8.766092634) (0.4999098002, 8.639396944) (0.5295011246, 8.516278434) (0.550188323, 8.436596868) (0.5662778028, 8.407745034) (0.5791035826, 8.381497522) (0.5903855586, 8.360780846) (0.5997335402, 8.347279044) (0.608103685, 8.34136958) (0.6159557148, 8.33774408) 
};
\addlegendentry{proposed, fittest}
\addplot[dashed, very thick, red, mark size = 1 pt, mark options = {solid, thin}]
coordinates {
(0.086663621422, 10.01506244) (0.107328993256, 10.01429518) (0.12876481919, 9.99547633) (0.148775796434, 9.98125433) (0.1675690556, 9.951686426) (0.19810646238, 9.87081989) (0.22270764192, 9.81771209) (0.24198303742, 9.721660666) (0.2581683898, 9.663241706) (0.27085223244, 9.60015803) (0.28255796086, 9.545322868) (0.29361976816, 9.515934882) (0.30168320924, 9.462724648) (0.31645890366, 9.405073922) (0.32990244492, 9.357283116) (0.34151160912, 9.316245414) (0.35131420306, 9.281419802) (0.3877354178, 9.160509462) (0.4143904902, 9.094173206) (0.4513921192, 9.012468558) (0.4788833498, 8.958628848) (0.5000962058, 8.929797982) (0.5173826246, 8.903679702) (0.5322993904, 8.886766112) (0.5455395396, 8.869230568) (0.557242117, 8.84995203) (0.5673326562, 8.841589188) 
};
\addlegendentry{edksp, fittest}
\addplot[dotted, very thick, red, mark size = 1 pt, mark options = {solid, thin}]
coordinates {
(0.086355280082, 10.010925302) (0.107100593176, 10.011369862) (0.12728235469, 9.990412122) (0.147537033634, 9.968836644) (0.1629804652, 9.93323624) (0.19373712338, 9.875995264) (0.21668548612, 9.798768228) (0.23710054442, 9.73829763) (0.252270374, 9.671937704) (0.26813430504, 9.630182102) (0.28014051126, 9.585742318) (0.29206624636, 9.545627904) (0.30137254424, 9.508684532) (0.31943920006, 9.446797876) (0.33380474712, 9.401698288) (0.34630187152, 9.351736194) (0.35840043126, 9.319481952) (0.4002842278, 9.173314876) (0.429382547, 9.094833136) (0.4706603972, 8.990522288) (0.5006188356, 8.919485022) (0.5231465794, 8.870307666) (0.5414597174, 8.831702526) (0.5572275366, 8.795588052) (0.5703908678, 8.764631878) (0.5823302018, 8.750305506) (0.5919572576, 8.72243013) 
};
\addlegendentry{yenksp, fittest}
\addplot[solid, mark size = 1 pt, mark options = {solid, thin}]
coordinates {
(0.086702837542, 10.015286926) (0.108298397536, 10.02090999) (0.12994017697, 10.015782612) (0.151802982934, 10.019827612) (0.17408250252, 10.016362884) (0.21837125666, 9.994524182) (0.26142735368, 9.95038011) (0.29739790528, 9.8717725) (0.32663134678, 9.77193141) (0.34507561722, 9.671075398) (0.36621243026, 9.570517556) (0.37911476176, 9.474110046) (0.39162461916, 9.387105058) (0.40909354352, 9.253720966) (0.4220238248, 9.15104597) (0.43338709998, 9.060039718) (0.44216048956, 8.9881306) (0.475437044, 8.756017602) (0.4966150342, 8.625939934) (0.5265456358, 8.492667874) (0.5467033538, 8.417407516) (0.5619558026, 8.368932518) (0.5754145046, 8.339547104) (0.5862118814, 8.312347974) (0.5958195578, 8.300424072) (0.60401357, 8.285937378) (0.6114761976, 8.270904924) 
};
\addlegendentry{proposed, first}
\addplot[dashed, mark size = 1 pt, mark options = {solid, thin}]
coordinates {
(0.086665838862, 10.012381294) (0.107918635556, 10.019153994) (0.12954746983, 9.99767762) (0.149540714834, 9.983933986) (0.168034899, 9.959317878) (0.20096933738, 9.891526352) (0.22513554252, 9.795187104) (0.24398380202, 9.720546118) (0.2600443686, 9.639193828) (0.27252377724, 9.597433776) (0.28406668726, 9.543988516) (0.29454112636, 9.490877384) (0.30349057124, 9.456182916) (0.31773460946, 9.387245558) (0.33089523912, 9.32112663) (0.34164904432, 9.27763767) (0.35105451726, 9.236476818) (0.3878739872, 9.107975784) (0.413792956, 9.037639822) (0.4509418896, 8.951858372) (0.4774484556, 8.895001376) (0.4978965204, 8.855984692) (0.5152316676, 8.832917904) (0.530098128, 8.8124153) (0.5432291, 8.787084626) (0.5539268922, 8.769261494) (0.564077479, 8.753902172) 
};
\addlegendentry{edksp, first}
\addplot[dotted, mark size = 1 pt, mark options = {solid, thin}]
coordinates {
(0.086515602382, 10.017756836) (0.107834904476, 10.01799479) (0.12740081217, 9.99324914) (0.146206386034, 9.978571966) (0.1639753978, 9.949810656) (0.19324621898, 9.869580412) (0.21717579772, 9.808397422) (0.23627081222, 9.723130816) (0.253481025, 9.684324286) (0.26743315404, 9.619740974) (0.28020710606, 9.57372338) (0.29107584596, 9.52712813) (0.30031955864, 9.498180424) (0.31905951866, 9.433298736) (0.33359707252, 9.374818672) (0.34584809692, 9.328360942) (0.35663481766, 9.290491494) (0.3985821714, 9.15472423) (0.4276507036, 9.05592441) (0.46878804, 8.951451922) (0.4973622534, 8.87560885) (0.5201774492, 8.828137084) (0.5381127192, 8.77652799) (0.5530367688, 8.74174315) (0.5654528078, 8.705240268) (0.577790159, 8.682785698) (0.5875427202, 8.659260792) 
};
\addlegendentry{yenksp, first}
\end{axis}
\end{tikzpicture}}\\[20pt]
  \end{tabular}
  \ref{legend}\\[10pt]
  \caption{Simulation results for Gabriel graphs.}
\end{figure*}
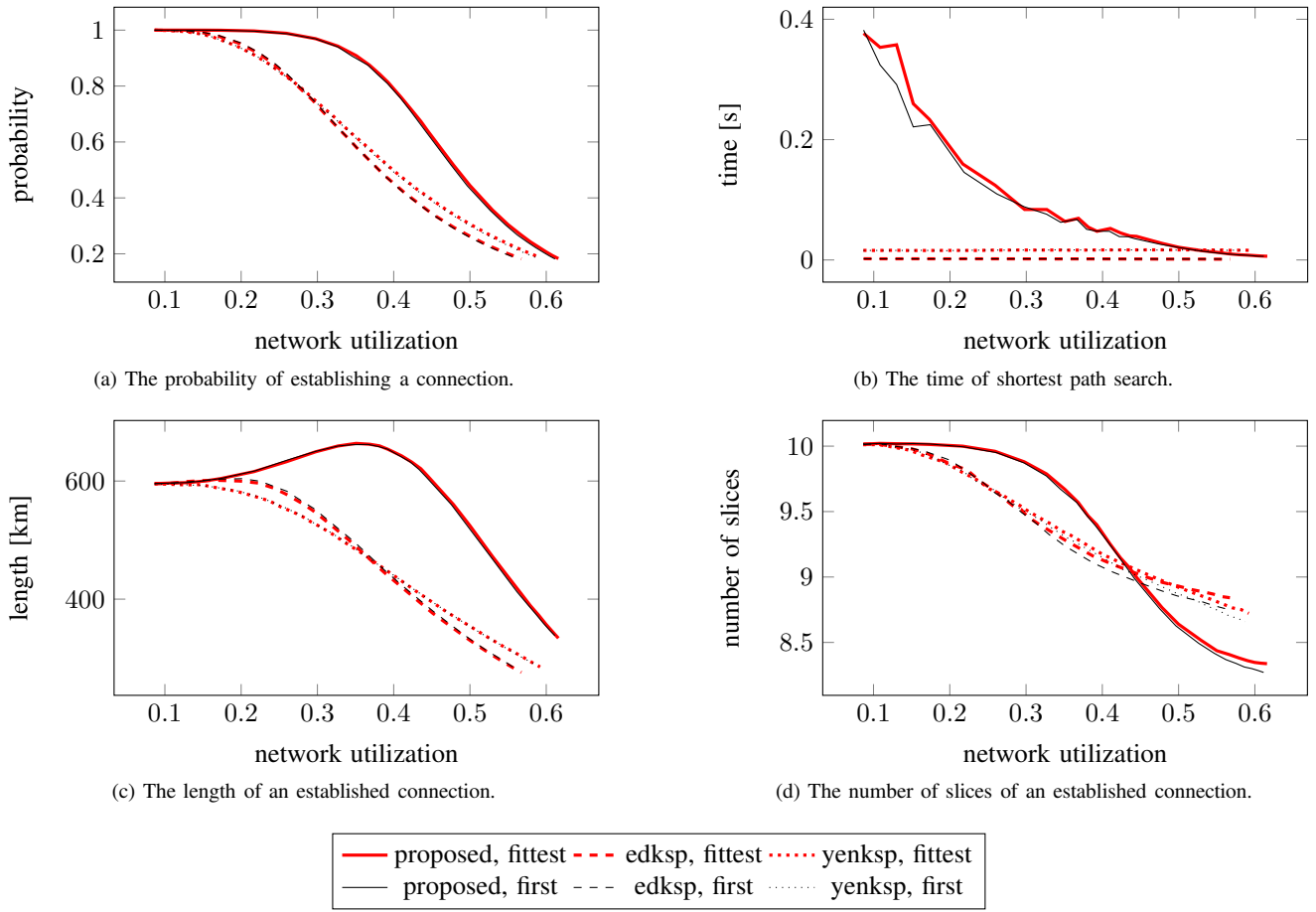

Figures \ref{f:pec}-\ref{f:nscec} show the simulation results for
routing with the proposed algorithm (solid curves), routing with the
edge-disjoint paths (dashed curves), and routing with the Yen $K$
shortest paths (dotted curves).  Routing was carried out for two
spectrum allocation policies: fittest (thick curves) and first (thin
curves).  Each figure has six curves for three routing types with two
spectrum allocation policies.  Each curve is plotted with 27 data
points for different values of $\lambda$.  Each data point represents
a sample mean, except the time of the shortest path search, which
represents a sample minimum.  Since the relative standard errors of
the sample means are below 1\%, the error bars would be too small to
plot.

Fig.~\ref{f:pec} shows the probability of establishing a connection as
a function of network utilization.  We are interested in how a network
performs for a given network state expressed by utilization.  The
proposed algorithm considerably outperforms the other routing types
for all network loads.  For the utilization of 30\%, the proposed
algorithm still has the probability of nearly 1, while for the other
two types, their probabilities drop to about 0.75.  For the load of
40\%, the probability for the proposed algorithm is almost twice as
large as that for the other two routing types.

Fig.~\ref{f:spsat} shows the time taken by a shortest path search as a
function of network utilization, regardless of whether the search was
successful or not.  The algorithms for finding edge-disjoint paths
and Yen $K$ shortest paths do not take into account the slices
available on edges, they do not depend on the utilization, and so
their times are constant.  Interestingly, the time taken by the
proposed algorithm decreases as the utilization increases, because
the search space gets narrower.

Fig.~\ref{f:lenec} shows the average length of an established
connection as a function of network utilization.  As the network is
utilized more, the average length of a connection increases, because
the proposed algorithm finds more circuitous paths, but still finds
them, while the other routing types fail.  The average length for all
routing types drops as utilization keeps increasing, since demands
with end nodes close to each other are more likely to be established.

Fig.~\ref{f:nscec} shows the average number of slices of an
established connection as a function of network utilization.  As
expected, the average number decreases as utilization increases,
because demands which require a smaller number of slices, are more
likely to be established.  At first glance, the proposed routing
performs better for network utilization below 45\%, and worse
otherwise than the other two routing types.  However, this is the
average number of slices, provided a connection was established.  As
the other two routing algorithms offer lower probabilities of
establishing a connection, they are more likely to establish short
connections, which are more likely to succeed demanding a larger
number of slices.  The proposed algorithm is more likely to establish
connections between pairs of distant nodes with a smaller number of
slices, thus lowering the average number of slices.

%%%%%%%%%%%%%%%%%%%%%%%%%%%%%%%%%%%%%%%%%%%%%%%%%%%%%%%%%%%%%%%%%%%%%%%%%%%

\section{Conclusion}
\label{conclusion}

We proposed the adaptation and the constriction of the Dijkstra
shortest path algorithm for finding shortest paths in elastic optical
networks.  The adaptation is a novel contribution, which takes into
account the spectrum continuity and contiguity constraints.

The routing and spectrum assignment problem is known to be
NP-complete, yet our algorithm has no difficulty finding a solution,
since, we speculate, limiting the path length narrows the search
space.  Limiting the path length is not a problem, since it is needed
in practice anyway.

Our extensive simulation studies show that the proposed algorithm
outperforms two other routing types frequently used in research on
elastic optical networks: routing along the edge-disjoint paths, and
routing along the Yen $K$ shortest paths.  The studies also show that
the algorithm can be used for routing in elastic optical networks of
large sizes, thus making the algorithm practical.

Future work could concentrate on removing the path length
constriction, and finding the stop condition for the algorithm, i.e.,
a condition to stop searching for a path, when it is known that no
path can be found.

\section*{Acknowledgments}

This work was supported by the postdoctoral fellowship number
DEC-2013/08/S/ST7/00576 from the Polish National Science Centre.  The
numerical results were obtained using PL-Grid, the Polish
supercomputing infrastructure.

\bibliography{all}

\end{document}